\documentclass[%
superscriptaddress,
rsi,
 amsmath,amssymb,
 reprint,%
]{revtex4-1}

\usepackage{graphicx}% Include figure files
\usepackage{dcolumn}% Align table columns on decimal point
\usepackage{bm}% bold math
\usepackage{upgreek}
\usepackage{siunitx}
\usepackage[utf8]{inputenc}
\usepackage[T1]{fontenc}
\usepackage{mathptmx}

\usepackage[usenames]{color}
\definecolor{amethyst}{rgb}{0.6, 0.4, 0.8}

\begin{document}

\preprint{AIP/123-QED}

\title[Active Optical Table Tilt Stabilization]{Active Optical Table Tilt Stabilization}

\author{Charles W. Lewandowski}
    \affiliation{Department of Physics, Montana State University, Bozeman, Montana 59717, USA}
\author{Tyler D. Knowles}
    \affiliation{Department of Mathematics, West Virginia University, Morgantown, West Virginia 26506, USA}
\author{Zachariah B. Etienne}
    \affiliation{Department of Physics and Astronomy, West Virginia University, Morgantown, West Virginia 26506, USA}
    \affiliation{Center for Gravitational Waves and Cosmology, West Virginia University, Chestnut Ridge Research Building, Morgantown, West Virginia 26505, USA}
\author{Brian D'Urso}
    \email{durso@montana.edu}
    \affiliation{Department of Physics, Montana State University, Bozeman, Montana 59717, USA}

\date{\today}

\begin{abstract}
We show that a simple modification to an optical table with pneumatic vibration isolation can be used to actively reduce the long term drift in the tilt of the table by nearly a factor of 1000. Without active stabilization, we measure a root-mean-square (RMS) tilt variation of $\SI{270}{\upmu rad}$ over three days. The active stabilization can be used to limit the tilt to $\SI{0.35}{\upmu rad}$ RMS over the same time period. This technique can be used to minimize drift in tilt-sensitive experiments.
\end{abstract}

\maketitle

Levitated optomechanics provide a unique system for studying fundamental physics~\cite{moore2018tests} as well as practical applications such as accelerometry~\cite{lewandowski2020high}. These systems, dominated by optical trapping~\cite{ashkin1971optical}, but also including magnetic traps~\cite{slezak2018cooling} are typically mounted on optical tables with pneumatic vibration isolation. While the relatively low stiffness of levitated systems makes them highly sensitive force or acceleration sensors~\cite{ranjit2016zeptonewton, lewandowski2020high}, it also makes them more sensitive to tilt than clamped resonators (e.g.~\cite{teufel2011sideband}). In this manuscript, we provide a solution to this issue by actively stabilizing the tilt of the table.

Since trapped particles are typically subject to a harmonic potential, the motion can be described as a simple harmonic oscillator with angular oscillation frequency $\omega=\sqrt{k/m}$, where $k$ is the spring constant of the trap and $m$ is the mass of the trapped particle. For a small tilt of the table $\Delta\theta$, the displacement of the particle $\Delta z$ from the unperturbed equilibrium position is 
\begin{equation}
    \Delta z \approx \frac{g}{\omega^2} \Delta \theta,
    \label{eq:shift}
\end{equation}
where $g$ is the acceleration due to gravity. For low frequency systems, such as those required for precision measurements of the Newtonian constant of gravitation~\cite{klahold2019precision}, the displacement can be significant enough to disrupt the measurement.

Pneumatic isolators are designed to minimize transmission of vibrations from the floor to the table, but may introduce susceptibility to long term drift in the tilt of the table due to changing environmental conditions. Changes in temperature or atmospheric pressure as well as typical laboratory activities may lead to unacceptable changes in the tilt of the table. In this study, the table used is $\SI{8}{feet}$ by $\SI{4}{feet}$ in the $x$ and $z$ directions, respectively, supported by four TMC Gimbal Piston Isolators.

\begin{figure}
    \includegraphics[width=1.0\columnwidth]{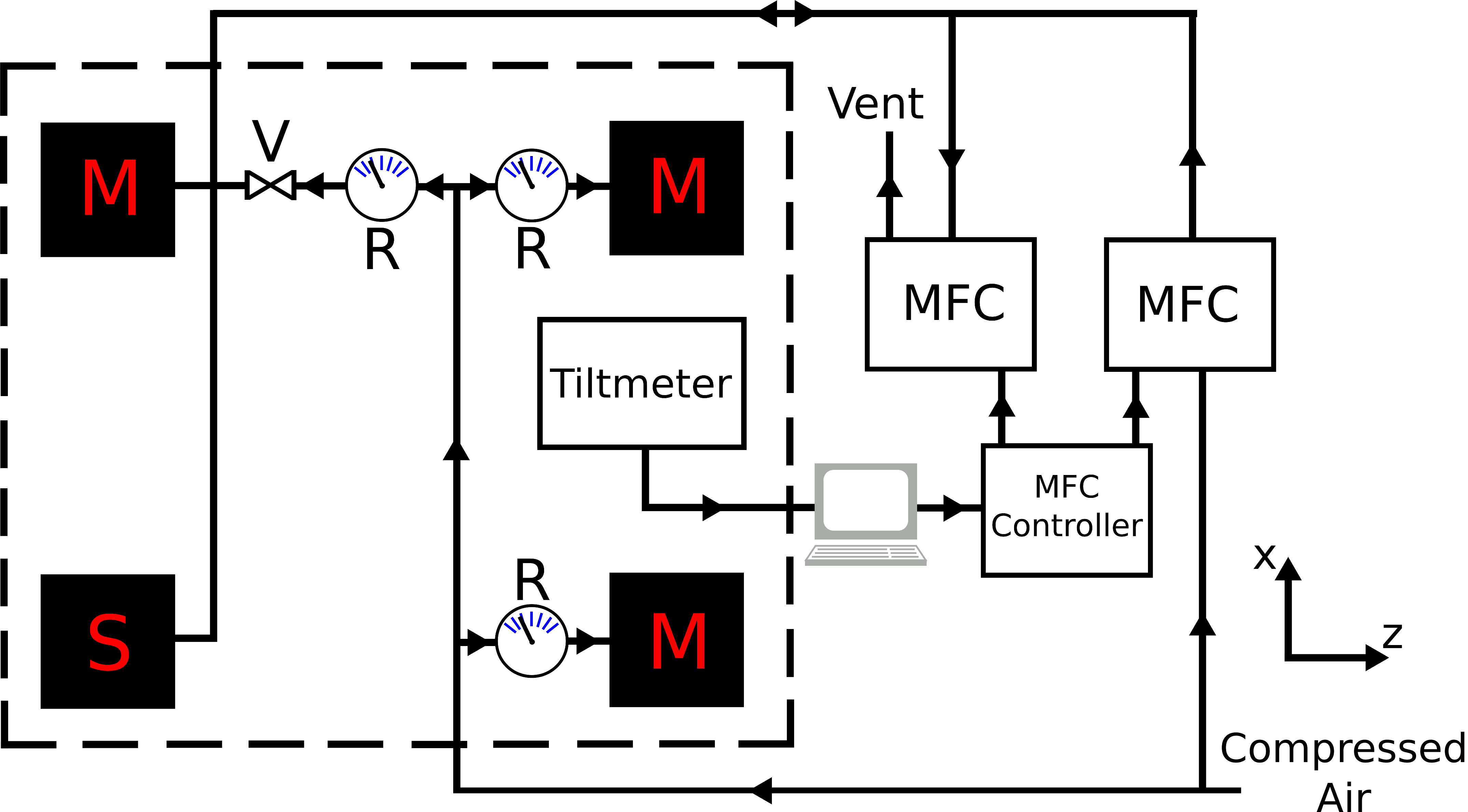}
    \caption{\label{fig:flow}System diagram for actively stabilizing the tilt of an optical table. Two mass flow controllers are regulated to add or remove compressed air from the master-slave leg pair, based on real-time readings from the tiltmeter. When stabilized, valve V is closed to the table regulator. The arrows indicate the direction of flow.}
\end{figure}

The original table support system consists of a regulator-valve system on each of the three master legs, labeled by R and M in Fig.~\ref{fig:flow}, respectively. The slave leg, labeled S, is tied to one master leg. In our case, the master-slave pair is parallel to the $x$ direction. The $z$ direction, which we stabilize, is parallel to the short side of the table. Under normal operation, the regulators add or remove compressed air from the isolators to keep the table approximately level.

To stabilize the tilt, a simple modification was made to the regulation of the isolators (see Fig.~\ref{fig:flow}). An air line inserted in the master-slave leg pair is immediately split into two lines and connected to two MKS mass flow controllers (MFCs) (MKS 1179C01312CR1BV). One MFC is oriented to allow gas flow towards the legs and the other is oriented to allow gas to flow out of the legs. The two MFCs are controlled by an MKS type 247 4-channel flow controller power supply and readout (MFC controller). An inline valve is inserted between the master-slave leg pair and its original regulator. The valve isolates the master-slave leg pair from the original regulator to prevent the regulator from competing with the flow from the MFCs when active stabilization is used.

\begin{figure*}[t!]
    \includegraphics[width=2.0\columnwidth]{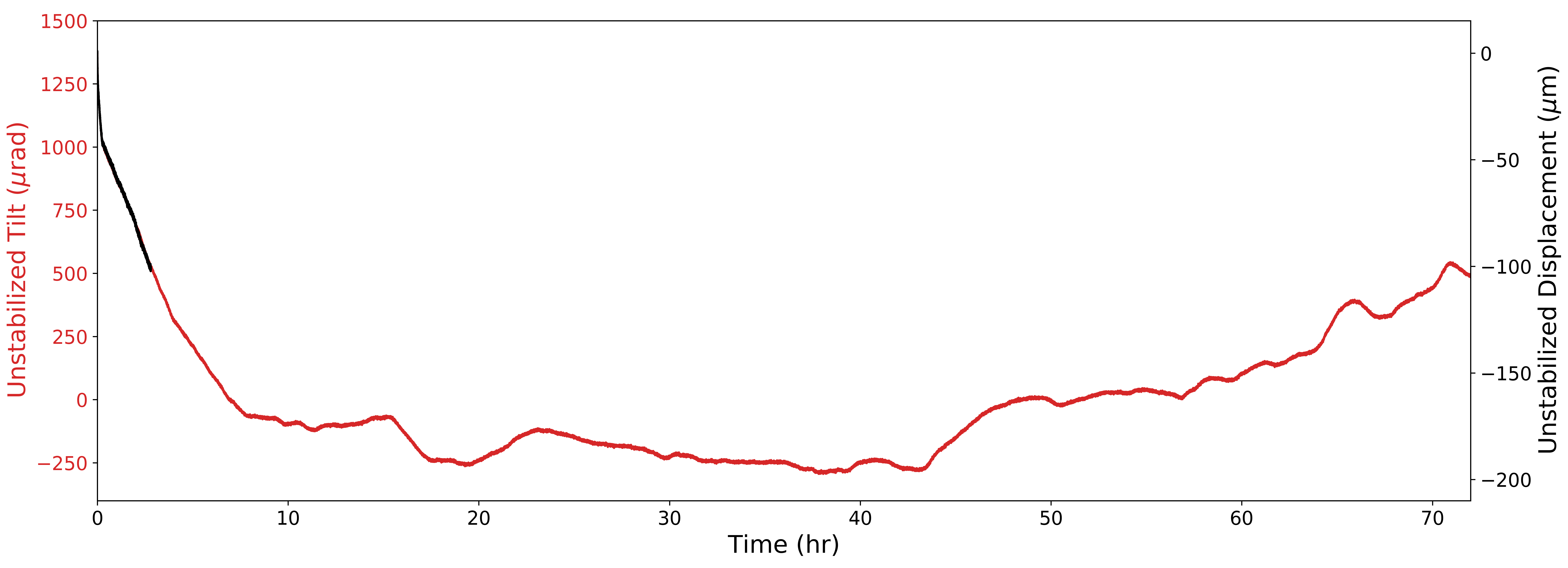}
    \caption{The relative tilt of the optical table (red) and displacement of a trapped particle (black) without active stabilization over three days. The particle displacement follows the shape of the relative tilt before the particle leaves the frame of the camera.}
    \label{fig:unstabilized}
\end{figure*}

A Jewell Instruments A603-C two-axis tiltmeter is placed on the optical table, centered in the $z$ direction, to monitor the tilt with a resolution of less than $\SI{2.5}{nrad}$ along two perpendicular axes. The high gain setting with angular range of $\pm 40$~\si{\upmu rad} and an optional $\SI{7}{s}$ output filter were used. In the initial setup of the meter, the two axes are oriented along the $x$ and $z$ directions of the table. The outputs are monitored on Keithley 2000 digital multimeters (DMM). The tiltmeter is adjusted by the worm-gear feet until the voltage output is near zero, corresponding to zero absolute tilt. Since we stabilize the relative tilt, it is not crucial to completely zero the tiltmeter.

A second tiltmeter is placed on the optical table, also centered in and aligned with the $z$ direction for coarse tilt detection when the A603-C does not provide enough angular range. The second is a Jewell Instruments LSOC-1Z fluid damped single-axis inclinometer with a $\pm 0.017$~\si{rad} range and $\SI{1}{\upmu rad}$ resolution. The single $\pm \SI{5}{V}$ output is monitored on a Keithley 2000 DMM.

\begin{figure*}[t!]
    \includegraphics[width=2.0\columnwidth]{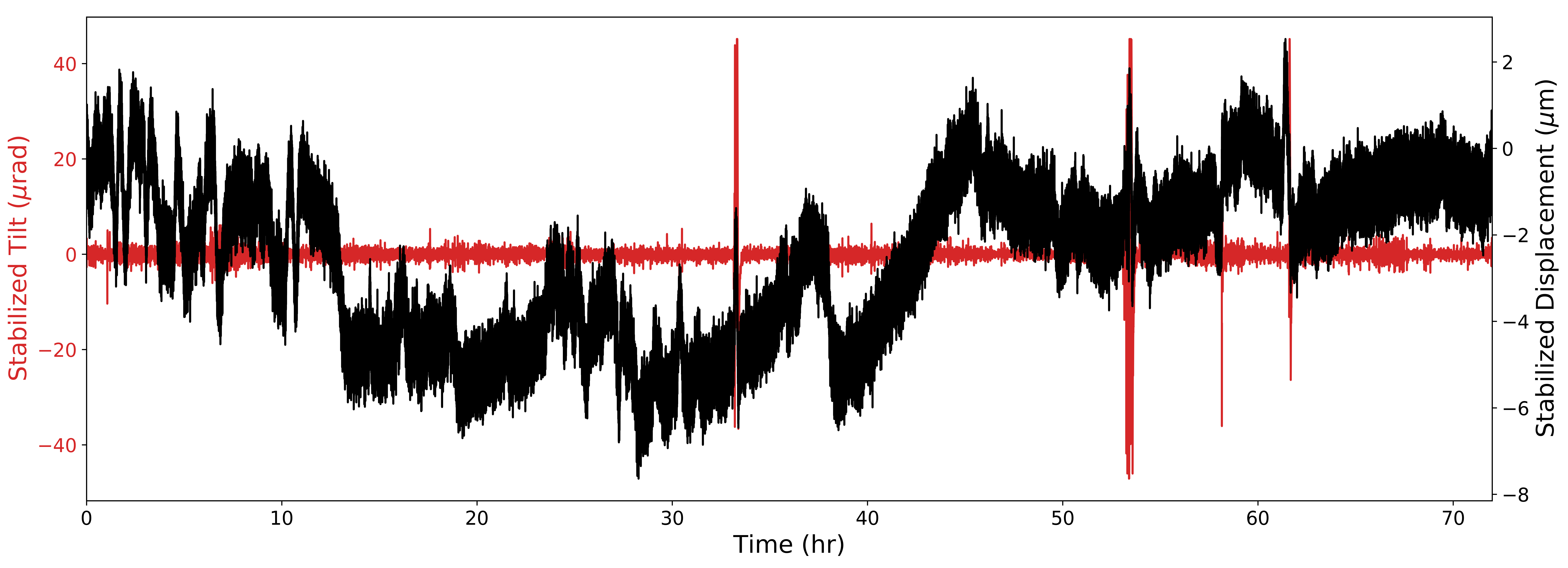}
    \caption{The relative tilt of the optical table (red) and displacement of a trapped particle (black) with active stabilization over three days.}
    \label{fig:stabilized}
\end{figure*}

The in-flow and out-flow rates are governed by a digital proportional-integral (PI) controller. For every new measurement, the difference between the measured tilt and target tilt, $\Delta_i$, is calculated. This error is multiplied by the proportional and integral constants, $P$ and $I$, respectively. The algorithm is described by 
\begin{equation}
    F_i = P \Delta_i + I\sum\limits_{n=0}^i\Delta_n
    \label{eq:PI}
\end{equation}
where $F_i$ is the feedback flow corresponding to the $i^{\textrm{th}}$ tilt measurement. To prevent excessive integral windup of the algorithm, the integral term is limited to $\pm\SI{216}{\upmu rad}$. To avoid an abrupt crossover from one mass flow controller to the other (when the sign of $F_i$ changes), the outputs were biased such that an equal (but nonzero) flow of gas passed through both MFCs when $F_i=0$.

The tilt of the table was monitored for $\SI{72}{hrs}$ with one tilt reading taken every second, without and with active stabilization. Data without active stabilization was taken after the table was settled under active stabilization. The feedback MFC controller was then turned off and the valve to the regulator on the master-slave leg pair was opened so the table was fully controlled by the table regulators. As shown in Fig.~\ref{fig:unstabilized}, the tilt immediately began drifting and settled over many hours, eventually drifting back in the opposite direction. The root-mean-square (RMS) tilt variation was $\SI{270}{\upmu rad}$, requiring the use of the coarse (LSOC-1Z) inclinometer . This not only demonstrates the drift of the tilt over time, but also how it can take the isolators to initially settle.

\begin{figure}[b!]
    \includegraphics[width=1.0\columnwidth]{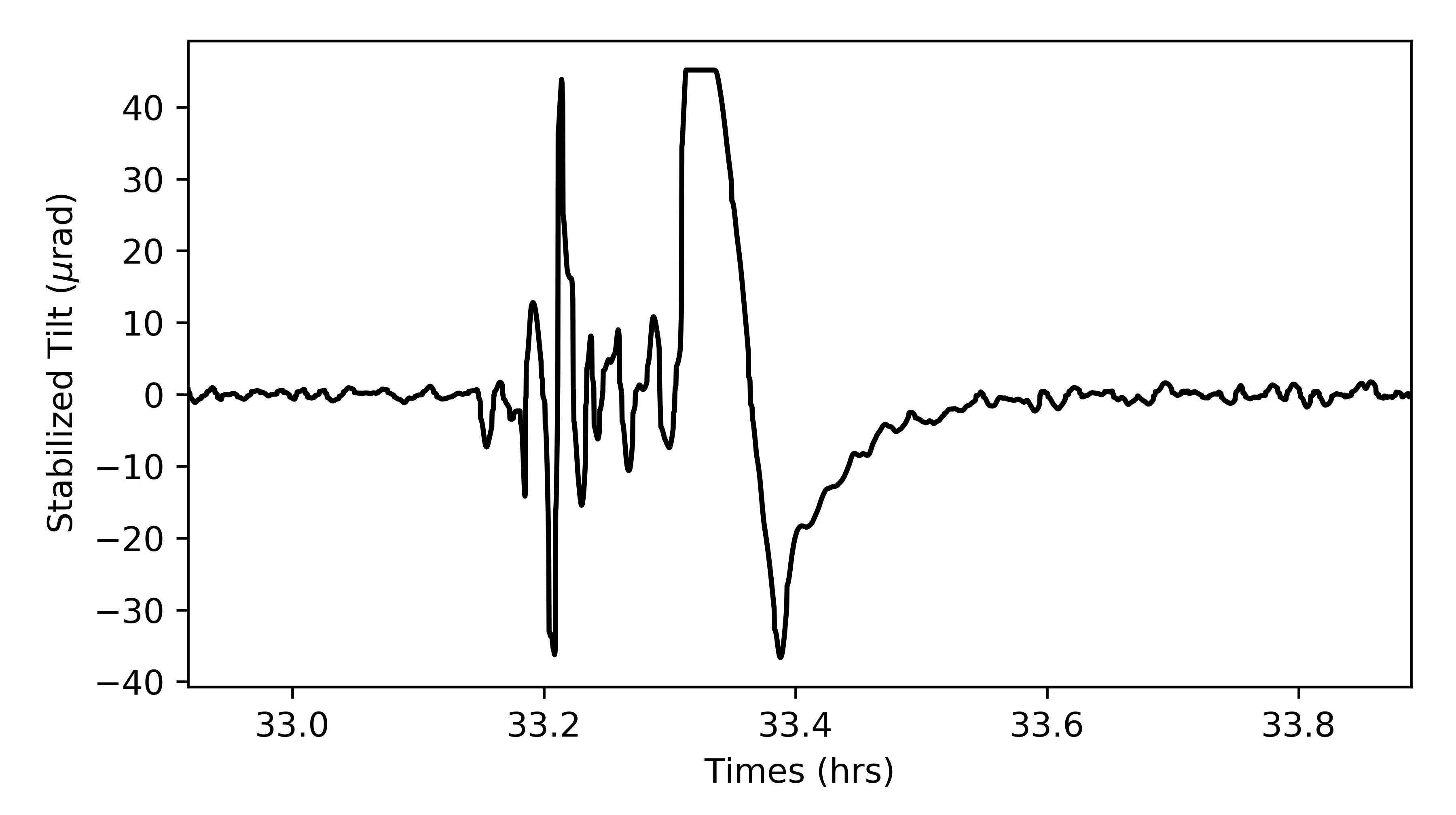}
    \caption{The stabilized data of an unknown event occurring at approximately $\SI{33.3}{hr}$. The relative tilt of the optical table exceeds the range of the A603-C tiltmeter. The entire event lasts $\sim\SI{1000}{s}$ before the stabilization corrects it.}
    \label{fig:close_up}
\end{figure}

To measure the tilt variation with active stabilization, the valve between the original regulator and master-slave leg pair was closed and the table was allowed to stabilize for several minutes before recording the data shown in Fig.~\ref{fig:stabilized}. There are several notable features in the stabilized data occurring at e.g. $\SI{33.3}{hrs}$. A detailed plot of this event is shown in Fig.~\ref{fig:close_up}. The cause of the disturbance is unknown but illustrates the performance of the stabilization. Before $\SI{33.1}{hrs}$, the optical table is well stabilized to within $\pm 1.2$~\si{\upmu rad}. After some overshoot, the disturbance is corrected within $\SI{1000}{s}$. Over the three day period, the RMS tilt variation was $\SI{2.8}{\upmu rad}$, including the large disturbances.

The effect of uncontrolled tilt on levitated optomechanics experiments can be dramatic. Particles in magneto-gravitational traps can have particularly low frequencies (we have demonstrated down to $\approx \SI{0.1}{Hz}$), making them particularly susceptible to tilt (see Eq.~\ref{eq:shift}). In this experiment, we trap $\SI{60}{\upmu m}$ diameter borosilicate microspheres in a trap that is $\SI{20}{mm}$ long in the $z$ direction. Air at atmospheric pressure is left in the trapping chamber to provide damping.

To measure the effects of tilt, images of a microsphere in a trap on the optical table were recorded with a CMOS camera while taking tilt data. The images were analyzed using the eigenframe method to obtain sub-pixel~\cite{lewandowski2020high} displacement of the particle. The analysis was performed offline after all frames were recorded. The recorded image is approximately $\SI{250}{\upmu m}$ horizontally and $\SI{125}{\upmu m}$ vertically, typical for our experiments.

The displacement of the particle during the tilt measurement without active stabilization is shown by the black curve in Fig.~\ref{fig:unstabilized}. The oscillation frequency of the particle is inferred by adjusting the scale appropriately for the particle displacement to match the relative tilt, according to Eq.~\ref{eq:shift}, and is found to be $\omega/(2\pi)=\SI{0.17}{Hz}$. In this data, the particle remains in frame until drifting approximately $\SI{125}{\upmu m}$ during the first $\SI{4}{hrs}$ due to the large tilt drift. The table never returned to the angle necessary for the particle to return to the frame. This behavior is unacceptable for the long-duration experiments we plan to pursue.

The particle displacement during the stabilized tilt measurement is shown by the black curve in Fig.~\ref{fig:stabilized}. The total range of the displacement is $\Delta z \approx \SI{10}{\upmu m}$, so the particle always remains near the center of the frame. During the four relatively large disturbances in tilt, the displacement also jumps. Besides these four instances, the displacement shows little correlation with the relative tilt, showing that the particle is not limited by the tilt stability. The origin of the particle motion is not clear, but may be due to changes in the charge of the particle over time (likely due to the presence of air in the chamber).

We have shown that with a simple modification of the vibration isolators of an optical table, its tilt can be actively stabilize to an RMS variation of $\SI{0.35}{\upmu rad}$. This is nearly three orders of magnitude less than without stabilization. The need for such stabilization is evident in the displacement measurements of a trapped microsphere on the table. Without stabilization, the particle leaves the frame of the camera; with stabilization, the particle moves by less than one diameter and the particle displacement drift is not limited by the tilt stability.

This work was supported by the NSF under awards PHY-1707789, PHY-1757005, PHY-1707678, PHY-1806596, OIA-1458952, and OIA-1003907; NASA under awards ISFM-80NSSC18K0538 and TCAN-80NSSC18K1488; a block gift from the II-VI Foundation; the state of West Virginia HEPC; and West Virginia University.  

% \nocite{*}
\bibliography{aipsamp}% Produces the bibliography via BibTeX.

\end{document}